# Manipulating Gradient Forces on Optical Tweezers using Bessel Beams

Leonardo A. Ambrosio[†], Michel Zamboni-Rached[††], and Hugo E. Hernández-Figueroa[†], *IEEE Senior Member*

[†]Faculdade de Engenharia Elétrica e de Computação (FEEC), Departamento de Microonda e Óptica (DMO), Universidade Estadual de Campinas (UNICAMP), Av. Albert Einstein, 400, 13083-970 – Campinas – SP
[††]Centro de Ciências Naturais e Humanas, Universidade Federal do ABC, Santo André – SP.

*Abstract* — In this paper, we show how one can change the stable equilibrium of a particle trapped into an optical tweezer by varying the intensity of superposed Bessel beams with different orders. The gradient forces acting on particles of different radii are determined, and the theoretical results indicates that it is possible to combine Bessel beams in such a way as to manipulate the particle into or out the centre of the beam by exploiting their ring-shaped intensity patterns, without any mechanical displacement of the lasers.

## I. INTRODUCTION

In the year of 1970, A. Ashkin performed the first experiments on optical trapping by the forces of radiation pressure [1,2], opening a new and exciting area of research in the physical sciences. Dielectric particles in the micron or submicron scale, much larger than the wavelength, could be trapped using the assumption that light photons possess momentum. If light is reflected or scattered from a surface, then the momentum of each photon is changed, which would account for a force, according to Newton's second law. Although the existence of the radiation pressure was proved at the beginning of the twentieth century [3], it took about seventy years to this light feature be used in manipulating dielectric particles and individual atoms by either two-beam traps [1] or levitation traps [4,5]. In this case, the stability was guaranteed by requiring gravity or electrostatic forces.

If the size of the particle is much larger than the wavelength, a second type of force can be described by Newton's second law. This happens when light traverses a dielectric particle and is refracted changing its direction and, consequently, its momentum. In this case, if the particle possesses an index of refraction greater than that of the surrounding media, the net force tends to pull it into the centre of the beam, radial to the direction of propagation. In 1986, Ashkin also proved that three-dimensional trapping of a dielectric particle was also possible for a single, highly focused beam [6], and since then, optical tweezers has found interesting applications in manipulating biological cells, such as trapping of viruses and bacteria [7], induced cell fusion [8], studies of chromosome movement [9] and cellular microscopy [10].

Recently, optical manipulation of particles using Bessel beams was demonstrated [11]. These beams are solutions of the Helmholtz equation, independent of the propagation direction and with a cross-sectional profile of a set of concentric rings [12,13]. Although these solutions are commonly referred as non-diffracting beams, this can be quite controversial. In fact, their central maxima are resistant to the diffractive spreading, but this happens in expense of the lateral energy that helps to continuously reconstruct them. Ideal Bessel beams cannot be realized experimentally, as it possesses infinite energy. But truncated solutions are possible within a predetermined range, after which the beam would spread and decay.

Due to their ring structure, Bessel beams have the ability of trapping simultaneously both high and low refractive index particles [14-16] in their bright rings and dark regions, respectively. Furthermore, because of the absence of a specific focus, they can trap several particles along their axis [17].

In this paper, we calculate radial forces on dielectric particles by superposing Bessel beams with different orders of Bessel functions. Theoretical analysis, followed by simulations, shows that it is possible to dislocate the points of stable equilibrium, manipulating particles by suitable changes in the intensities of such beams.

## II. THEORETICAL ANALYSIS

Suppose an incident beam impinges on a small homogeneous non-absorbing dielectric particle, of radius $a$ much smaller than the wavelength $\lambda$, and with an index of refraction $n_d$. The total force per unit volume exerted on this particle can be written as [18]

$$\mathbf{f}(\mathbf{r}) = \frac{\alpha n_m}{2c} \nabla I_m(\mathbf{r}) + \alpha n_m \frac{\Delta m(\mathbf{r})}{\Delta t}, \qquad (1)$$

where $c$ is the speed of light in vacuum, $n_m$ is the index of refraction of the surrounding medium, $\Delta m(\mathbf{r})/\Delta t$ is the change of momentum density per unit time, $\alpha = 3(m^2-1)/(m^2+2)$ is the polarizability of the particle with $m = n_d/n_m$ and $I_m(\mathbf{r})$ is the intensity pattern of the incident beam. If one considers that the Rayleigh-Gans approximation is satisfied, i.e., $|m-1| \ll 1$ and $4\pi a \ll \lambda/|m-1|$, the second term on the right side of (1), corresponding to light scattering by the particle, can be neglected.

If we have an incident linearly polarized $n$th-order Bessel beam, for example, on the form

$$E(r,\phi,z) = E_0 J_n(k_t r) e^{im\phi} e^{ik_z z}, \quad (2)$$

with $k_z$ being the longitudinal wave number associated to the propagation along the optical axis $z$, and $\phi$ the azimuthal angle with $m$ an integer, its intensity is proportional to $J_n^2(k_t r)$, $r$ being the distance from de centre of the beam to the centre of the particle, and $k_t$ the transverse wave number of the beam. Performing the integration of (1) over the volume of the particle (hereafter assumed as a sphere), and after some manipulations we find, for the total field:

$$F(\mathbf{r}) \propto I_{max,n} \int_0^\pi \int_0^{2\pi} \int_0^\pi J_{2n}(2f(r,\theta,\varphi)\sin\xi)\sin^2\theta\cos\varphi\, d\xi\, d\varphi\, d\theta. \quad (3)$$

In (3), $I_{max,n}$ is the maximum value of the nth-order Bessel beam intensity, $\theta$ and $\varphi$ are spherical coordinates referred to a system centered at the centre of the particle, $\xi$ is an auxiliary variable for the integration [19] and $f(r,\theta,\varphi)$ equals

$$f(r,\theta,\varphi) = \left[(k_t r)^2 + (k_t a)^2 \sin^2\theta + 2(k_t r)(k_t a)\sin\theta\cos\varphi\right]^{1/2}. \quad (4)$$

Note that the total field is proportional to the intensity's highest value, which could occur at the centre of the beam for $n = 0$, or at some distance $r$ from it for $n \neq 0$. We could go further, and calculate the total force when two ore more Bessel beams are presented by using superposition in (2). Obviously, each Bessel beam would contribute according to its intensity, and interesting situations may arise if we can control, in real time, such contributions. It is worthy to say that, in evaluating (2), negative forces pull the particle radially to the centre of the beam, whereas positive forces push it away. As it is known, points of stable equilibrium would be possible every time we have, besides $F(\mathbf{r}) = 0$ at a specific point, restoring forces close to that point, bringing the particle back to its initial position.

## II. SIMULATION RESULTS

The simplest case that we can analyze happens when two Bessel beams of $n = 0$ and $n = 1$ are superposed, and their axial axis are coincident. Such simplicity is not only algebraic, but includes the experimental realization, because zero-order Bessel beams can be easily generated by annular apertures, optical elements as axicons or holography, while high-order Bessel beams are achieved with crystals that present some anisotropy, as biaxial crystals, for example [20]. Fig. 1 shows the results for a small-size sphere of $k_t a = 0.10$ when $I_{max,1}$ is normalized to 0.9, 0.60, 0.3 and 0, with corresponding normalized $I_{max,0} = 0$, 0.3, 0.60 and 0.9 $I_{max,1}$. The forces are in solid lines, and intensities are dashed. The horizontal axis is proportional to the distance from the particle to the centre of the beam. It can be seen that the points of stable equilibrium, marked with arrows, are shifted to different positions. Therefore, a particle initially at such points tends to be shifted as well, being dislocated toward the centre of the beam. This situation would be reversed if we keep $I_{max,0}$ fixed and normalized while varying $I_{max,1}$, implying on a shift outwards.

In Fig. 1(c), we can notice a transition along this process. This happens because the zero-order Bessel beam intensity becomes higher than that of the first-order beam. This causes the fastest shifts toward the optical axis of these beams, before and after which, the shift slows down.

Naturally, this situation can be more complicated in practice. The azimuthal displacement is not predicted in this case, for real Bessel beams may have imperfections in their concentric simetric rings, causing the particle to rotate about the beam axis. It must also be emphasized that the velocity in which the intensity is varied my cause the particle to escape from the trap, due to the presence of hydrodynamic forces.

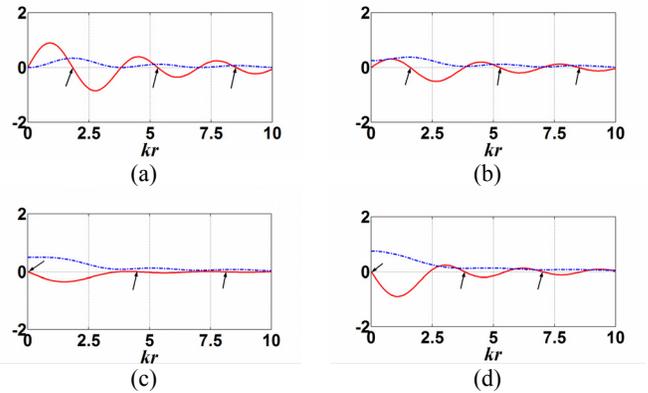

Fig. 1.   Intensity (dashed) and force (solid) calculations with $ka = 0.1$ for (a) . $I_{max,1} = 0.9$ and $I_{max,0} = 0$, (b) $I_{max,1} = 0.6$ and $I_{max,0} = 0.3$, (c) $I_{max,1} = 0.3$ and $I_{max,0} = 0.6$ and (d) $I_{max,1} = 0$ and $I_{max,0} = 0.9$. The points of stable equilibrium are shifted towards the origin.

To visualize the influence of the particle size to the resulting optical force, let us suppose $k_t a = 1.5$ and 3.5. The results for these situations are shown in Figs. 2 and 3, respectively. The same intensity variation for $I_{max,0}$ and $I_{max,1}$ was assumed. The stable equilibrium points are still seing for $k_t a = 1.5$, but disappear almost completely for $k_t a = 3.5$. This suggests that larger particles would not be shifted, but rather, they would tend to be stretched in an unusual way, with some parts of it experiencing a stronger force. Again, the situation is reversed if we interchange $I_{max,0}$ and $I_{max,1}$.

The linear variation of the intensities would be restricted, in practice, to the experimental setup available. If one were to control the intensity of only one Bessel beam, keeping the other fixed, independent laser beams would have to be used. A polarizer, for example, does not permit such flexibilities.

Finally, it should be pointed that the situation in Fig. 1 coincides with similar results in the literature [20], except for the fact that here the reference for positive radial force is assumed to be the radial versor pointing away from the beam centre, not towards it.

High-order Bessel beams could also be combined, with similar results. But in this case, there would be no stable equilibrium points around the origin, as these beams possess dark regions of intensity at this region. But as an illustrative example, for two Bessel beams of first- and second-order with $k_t a$ = 0.1, it can be seen in Fig. 4 that the displacement of the particle is smoothed, being shifted without a great variation in its velocity.

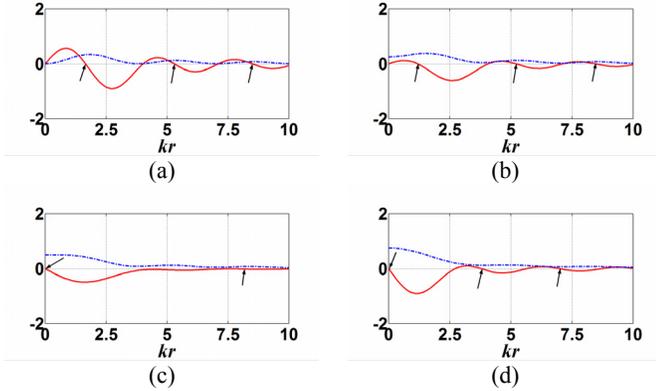

Fig. 2. Intensity (dashed) and force (solid) calculations with $k_t a$ = 1.5. The same intensities for each Bessel beam as for Fig. 1 were considered.

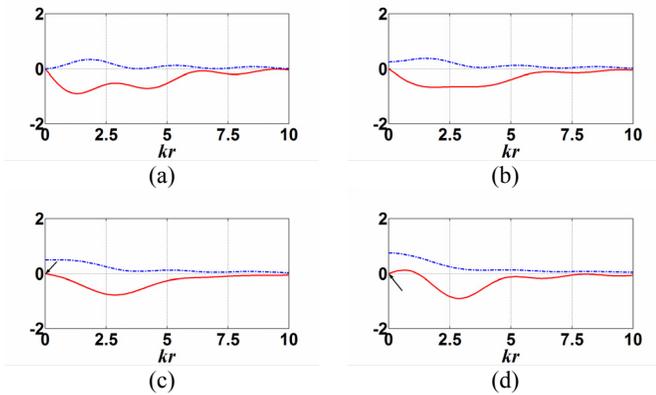

Fig. 3. Intensity (dashed) and force (solid) calculations with $k_t a$ = 3.5. The intensities were considered as before. Note the almost complete absence of equilibrium points. For the size assumed, it may be possible to stretch the particle.

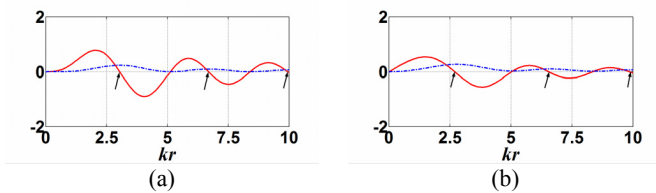

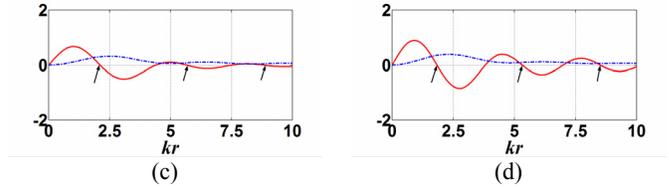

Fig. 4. Two Bessel beams of first- and second-order, with $k_t a$ = 0.1. The combination of high-order Bessel beams can shift the particle without annulling the points of stable equilibrium, as noticed before. This is because these beams all have dark regions of intensity around their optical axis, and their varying combinations results in a smoothed transition when the peaks of intensity are interchanged.

### III. CONCLUSIONS

A simple method for calculating the total force exerted by superposed Bessel beams on dielectric spheres of different radii, but restricted to the Rayleigh-Gans regime, or even in geometrical optics was presented. This force was shown to be dependent of the intensity of each beam individually, and this allowed for the computation of possible points of stable equilibrium.

The results suggest that optical manipulation can be realized in the radial direction with multiple Bessel beams, just by controlling its intensities. Depending on the size of the particles, they may shift toward or away from the axial axis of the beams, which were considered coincident. Larger particles can experience different force intensities along their radial length. This suggests a further investigation so as to understand at what extent one could perform elasticity measurements on dielectric particle including, for example, biological cells and molecules.

This work was supported by FAPESP – *Fundação de Amparo à Pesquisa e ao Ensino do Estado de São Paulo*, under contracts 2005/54265-9 (Ph D grant) and 2005/51689-2 (CePOF, Optics and Photonics Research Center).